# Static Deadlock Detection in MPI Synchronization Communication

Liao Ming-Xue, He Xiao-Xin, Fan Zhi-Hua
*Institute of Software, the Chinese Academy of Sciences, 100080, China*
liaomingxue@sohu.com

**Abstract**

*It is very common to use dynamic methods to detect deadlocks in MPI programs for the reason that static methods have some restrictions. To guarantee high reliability of some important MPI-based application software, a model of MPI synchronization communication is abstracted and a type of static method is devised to examine deadlocks in such modes. The model has three forms with different complexity: sequential model, single-loop model and nested-loop model. Sequential model is a base for all models. Single-loop model must be treated with a special type of equation group and nested-loop model extends the methods for the other two models. A standard Java-based software framework originated from these methods is constructed for determining whether MPI programs are free from synchronization communication deadlocks. Our practice shows the software framework is better than those tools using dynamic methods because it can dig out all synchronization communication deadlocks before an MPI-based program goes into running.*

## 1. Introduction

Deadlock is a very common problem in software designing and it may cause a running software program to break down. Deadlock in big application may even result into great loss, for example, the deadlock happened in the control software on NASA's Pathfinder landed on Mars 0. In 1971 Coffman addressed three strategies to process deadlocks: deadlock prevention, deadlock avoidance and deadlock detection and recovery 0. Deadlock prevention and avoidance have many deficiencies so that they are often used in systems which require high reliability 000. MPI 0 is a library specification for message-passing, proposed as a standard by a broadly based committee of vendors and users. It was designed for high performance on both massively parallel machines and on workstation clusters, however, it is very difficult to debug software programs based on it 00. Currently there are a few tools based on dynamic methods to debug errors in MPI programs, especially to detect deadlocks in them 0000. Both 0 and 0 need to insert some hand-shake codes into user's source programs to gather status of nodes to determine a deadlock. W. Haque utilizes MPI Profiling interface to intercept all MPI routine calls in order to check deadlocks 0. 0 is to find kinds of errors in MPI programs and uses MPI Profiling interface too, however, its interest covers not only deadlocks but also other kinds of errors.

However, these dynamic deadlock detection methods have a deadly deficiency that we can do nothing but suffering oncoming disaster when the deadlock happens. Systems requiring high reliability can not suffer this deficiency. For example, if an on-satellite cluster for monitoring rural flood or crops breaks down from a deadlock the life and economic loss will be innumerous. Therefore static methods are necessary to be developed to serve in such environments.

This paper introduces a static method to detect deadlocks in MPI synchronization communication. This static method is totally different from a static method in 0 which is based on techniques of searching finite state machine.

The second section defines a model of MPI synchronization communication. The model takes three different forms which are sequential model (S-Model), single loop model (L0) and nested loop model (L2). S-Model is the most basic models. We need to transform L0 and L2 into S-Model at appropriate time in order to detect deadlocks in them. To detect deadlocks in L0 involves a special type of equation group called ratio equation group. Algorithm for L2 deadlock detection is a combination of methods for L0 and S-Model.

Section 3 demonstrates how to examine deadlocks in a sequential model and section 4 and 5 follows to discuss L0 and L2. Section 6 gives an overview of our software framework for determining whether MPI programs are free from synchronization communication deadlocks and concludes this paper.

## 2. Modeling MPI programs

The MPI program studied in this paper is in Fig 1.

*program*::=*node - program*+ /*MPI program consists of programs running on each nodes*/
*node - program*::=<*nodeID*,*statements*>/*Node program includes a node ID and statements*/
*statements*::=*statement*+/*Statements are a non-null set of statement*/
*statement*::= *sequence - statement*|*for - statement* /*A statement is either a sequential one or a loop one*/
*sequence - statement*::=*send*|*receive*/*A sequential statement is either a MPI synchronizaiton send API call or a receive one*/
*for - statement*::=for(*n*){*statements*}/*A loop statement includes loop times n and statements*/

**Fig 1 MPI Synchronization Communication Model *L2***

Programs taking the form in Fig 1 are permitted to contain multi-layer nested loops but have no conditional statements included. Such programs are in a model called ***L2***. To explain why conditional statements are not covered in ***L2*** let us see an MPI program (1):

> *Process(machine)0*      *Process(machine)1*
> if(condition 1)          if(condition 2)                (1)
> send Msg *a* To *Process1*   recv Msg *a* From *Process0*

To detect deadlocks in (1) requires some dynamic techniques which are not included in this paper.

## 3. Sequential model

Model ***L2*** is called a sequential model (***S-Model***) if it has no loop statements. Program (2) is an example of ***S-Model***:

> *P0*                     *P1*
> send Msg *a* To *P2*     recv Msg *b* From *P0*
> send Msg *b* To *P1*     send Msg *c* From *P2*
>                                                        (2)
> *P2*
> recv Msg *c* From *P1*
> recv Msg *a* From *P0*

To detect deadlocks in (2), the first step usually is to build its ***Message Dependence Graph*** (***MDG***). The ***MDG*** of (2) is shown in Fig. 2:

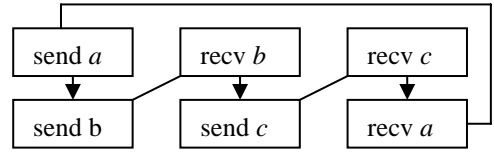

Fig. 2 ***MDG*** of (2)

This figure represents a directed graph. Line without an arrow is bidirectional. A circle with length greater than 2 is "send *a* → send *b* → recv *b* → send *c* → recv *c* → recv *a* → send *a*".

***MDG*** in Fig. 1 contains a circle which length is 6 greater than 2 so that we declare a deadlock in this ***MDG***. As a result program (2) has deadlocks. The deadlock of MDG of (2) indicates a situation: all three processes can not forward one step while waiting for other processes forms a circle. Theorem 1 discloses this fact:

**Theorem 1** An ***S-Model*** has no deadlock if and only if its ***MDG*** has no circle with length greater than 2.

Circle detection is often used to check deadlocks. Our ***MDG*** includes a special circle between a pair of matching messages and this special circle (which length is 2) does not mean a deadlock so that theorem 1 excludes this case. In fact, an ***MDG*** contains all temporal relationship among all messages.

How to build ***S-Model***'s ***MDG*** is not included in this paper. Moreover, this paper does not explain why theorem 1 holds. Strict proof of theorem 1 needs many definitions. Intuitively the theorem is correct.

However, it is costly to directly use theorem 1 to check an ***S-Model***'s ***MDG*** because searching for a length-more-than-2 circle is not very easy and building an ***MDG*** is costly too. Checking ***MDG*** is very frequent in our software framework for detecting MPI deadlocks so that we developed a very efficient algorithm to find deadlocks in MPI ***S-Model*** instead of finding circles in ***MDG***. Below is a brief introduction to the algorithm.

Firstly, the sequential model is mapped into a set of character strings and its deadlock detection problem is translated into an equivalent multi-queue string matching problem. The following step is an infinite loop until all queues become empty or any queues can not be updated again. If the loop ends when any queues can not be updated again, we declare a deadlock. Each time the loop starts, we update two queues by removing matching messages whenever a pair of matching message is found in the two queues. The time and space complexity of the algorithm is O(n) where n is the amount of message in model.

## 4. L0 Model

Model *L2* is called *L0* if it has no nested loop statements. Program (3) is an example of *L0*:

*P0*
for(∞)
  send Msg *a* To   *P1*
  send Msg *c* To   *P2*
  recv Msg *b* From *P1*
end-for

*P1*
for(∞)
  recv Msg *a* From *P0*,   send Msg *b* To   *P0*   (3)
  recv Msg *a* From *P0*,   recv Msg *d* From *P2*
  send Msg *b* To   *P0*,   recv Msg *d* From *P2*
end-for

*P2*
for(∞)
  recv Msg *c* From *P0*
  send Msg *d* To   *P1*
end-for

The process of detecting deadlocks in program like (3) has 3 steps. One is to build a ***Ratio Equation Group (REG)*** of the program and try to find its solution; if no solution we report a deadlock in the program. The second is check ***ratio consistency*** of the program; if inconsistent we assert a deadlock. The last step is to check whether an ***S-Model*** sliced from the program has deadlocks; if the ***S-Model*** has deadlocks we declare a deadlock in the program.

The beginning two steps are to check whether there is a type of matching message (e.g. message *a* in (3)) which occurrences are not absolutely equivalent in two node programs and are to find an optimal way to reduce loop times. The third step is then to check whether all messages are arranged in an order that results into a deadlock.

**The first step** From the perspective of message *a*, there are one occurrence of message *a* in program 0 (with the loop times ignored) and two occurrences in program 1 so that the ratio of *p0* to *p1* is 1:2. All such ratios are listed in (4) which is a ***Ratio Equation Group***:

$$\begin{cases} p0:p1 = 1:2 \\ p0:p2 = 1:1 \\ p0:p1 = 1:2 \\ p1:p2 = 2:1 \end{cases} \quad (4)$$

We have developed a Java-based library for Ratio Equation Group which has an $O(n)$ time-space complexity where $n$ is the number of variables in group. Using this library we get the solution (5) to (4)

$$p0: p1: p2 = 1:2:1 \quad (5)$$

**The second step** We use theorem 2 to check ratio consistency.

**Theorem 2** A *L0* is ***ratio consistent*** if and only if

$$(\forall i, j)(p_i * t_i = p_j * t_j)$$

holds where
  *i* and *j* are non-negative integers to identify different processes (nodes);
  $p_i$ and $p_j$ are different processes' values in solution to *L0*'s ***ratio equation group***;
  $t_i$ and $t_j$ are loop times of process *i* and process *j* ; if loop times are infinity then the corresponding value of *t* is reset to zero.

According to theorem 2 and

$$\begin{aligned} p_0: p_1: p_2 &= 1:2:1 \\ t_0: t_1: t_2 &= \infty:\infty:\infty \end{aligned} \quad (6)$$

we assert the *L0* program (3) is ***ratio consistent***.

**The last step** This step is to slice an *S-Model* from *L0* and check the *S-Model* instead of checking the *L0*. Firstly we need to find the least common multiple (*LCM*) of all values of variables in solution to *L0*. The *LCM* of (5) is 2. Then slice each node-program of *L0* to form an *S-Model* of the *L0*. In other words we are to reset loop times of each node-program according to

$$\text{loop times of } node\text{-}program\ i = LCM / p_i \quad (7)$$

After being reset loop times according to (7) the program (3) is changed to

```
P0                              P2
for(2)                          for(2)
  send Msg a To   P1              recv Msg c From P0
  send Msg c To   P2              send Msg d To   P1
  recv Msg b From P1            end-for
end-for
```
(8)

```
P1
for(1)
  recv Msg a From P0,  send Msg b To   P0
  recv Msg a From P0,  recv Msg d From P2
  send Msg b To   P0,  recv Msg d From P2
end-for
```

All loop times in (8) are small enough so that we can easily give an equivalent *S-Model* of (8):

```
P0                              P2
  send Msg a To   P1              recv Msg c From P0
  send Msg c To   P2              send Msg d To   P1
  recv Msg b From P1              recv Msg c From P0
  send Msg a To   P1              send Msg d To   P1
  send Msg c To   P2
  recv Msg b From P1
```
(9)

```
P1
  recv Msg a From P0,  send Msg b To   P0
  recv Msg a From P0,  recv Msg d From P2
  send Msg b To   P0,  recv Msg d From P2
```

*S-Model* (9) is the equivalent *S-Model* of (8) and (3). Now we can check this *S-Model* (9) using theorem 1.

Summarizing the three steps we have theorem 3:

**Theorem 3** A *L0* is free from deadlocks if and only if it is *ratio consistent* and its equivalent *S-Model* is free from deadlocks. (Proof is not given here)

Using theorem 1 we can assert (9) free from deadlocks, so program (3) are free from deadlocks too by theorem 3.

## 5. L2 Model and deadlock detection

Let us see a somewhat complex MPI *L2* program (10).

```
P0
for(∞)
  for(2)
    send Msg a To P1,  send Msg c To  P1
  end-for
  for(4)
    send Msg b To P2
  end-for
  send Msg a To P1
  send Msg c To P1
end-for
```

```
P1                              P2
for(∞)                          for(∞)
  for(3)                          for(4)
    recv Msg a From P0              recv Msg b From P0
    recv Msg c From P0            end-for
  end-for                         recv Msg d From P1
  send Msg d To P2              end-for
end-for
```
(10)

For convenience, firstly we map (10) into a character string set.

$$P0 \to ((ac)^2 b^4 ac)^\infty$$
$$P1 \to ((ac)^3 d)^\infty \quad (11)$$
$$P2 \to (b^4)^\infty$$

Next, we must reduce the string set to its simplest form. A string such as $(ac)^2$ and $b^4$ is called *Power*. We state that a string set is in the simplest form if and only any string in the set does not contain powers in forms like $(x^{p_1})^{p_2}$ or $x^{p_1}(xy)^{p_2}$. The former form $(x^{p_1})^{p_2}$ must be changed to $x^{p_1 \cdot p_2}$, and this transformation is called *Power Reduction*. The latter $x^{p_1}(xy)^{p_2}$ must be changed to $x^{p_1+1}y(xy)^{p_2-1}$, and this transformation is called *Left Prefix Reduction*.

A *Ratio Equation Group* can also be build for a *L2* and the solution to the *REG* of (11) is

$$p0: p1: p2 = 1:1:1$$

So (11) can be reduced to (12) according to methods used for *L0*.

$$P0: (ac)^2 b^4 ac$$
$$P1: (ac)^3 d \quad (12)$$
$$P2: b^4$$

Now each node-program in (12) can be seen as some continued **Powers**. We define **FPP (First Power Pool)** as a set of powers consisting of the first most outer power of each node-program. The **FPP** of (11) is (11) itself. The **FPP** of (12) is

$$P0:(ac)^2$$
$$P1:(ac)^3 \quad (13)$$
$$P2:b^4$$

Note that currently (13) becomes a **L0** model. Therefore, we can build the **REG** of (7) and its solution is

$$p_0:p_1 = 1:1, \ p_2 = 1 \quad (14)$$

Solution (14) is different from other solutions because node-program *P0* and *P1* is related and node-program *P2* is irrelevant to them.

If we follow the method to detect deadlocks in **L0**, then (13) is found a deadlock. The fact is just the opposite. Based on (14), we can expand (13) (this method similar to the method in section 4 to slice a **L0**, and we say programs like (13) is **expansible** if its **REG** has a solution) to

$$P0:(ac)^2$$
$$P1:(ac)^2 ac \quad (15)$$
$$P2:b^4$$

The **FPP** of (15) is

$$P0:(ac)^2$$
$$P1:(ac)^2 \quad (16)$$
$$P2:b^4$$

Then we can assert (16) is **reducible** because *P0* and *P1* form a **L0** which is free from deadlock and we remove the reducible part from (12). A new **FPP** of (12) becomes

$$P0:b^4$$
$$P1:ac \quad (17)$$
$$P2:b^4$$

Repeat this reduction process we eventually find that **FPP** of (12) becomes empty. Until now, we determine that program (11) is free from deadlocks.

If (13) is not a **L0**, for example, suppose (13) to be

$$P0:(ac^5)^2$$
$$P1:(ac^5)^3 \quad (18)$$
$$P2:b^4$$

Solution to (18) remains the same as (14). We can operate on (18) following the process of (13) to (15) and the result is the same.

From the above process of deadlock detection in (10), we can devise an algorithm (19) for detecting deadlocks in **L2**.

```
checkDeadlock()
input: L2
reduce L2 to its simplest form
while(FPP of L2 is not empty)
  reduced ← false
  if(FPP is not empty)
    for each related set rs in FPP
      if(rs is reducible)   reduce rs, reduced ← true
      else-if(rs is expansible)                           (19)
        expand rs according to method in section 4
        reduced ← true
      end-else-if
    end-if
    end-for
  end-if
  if(¬ reduced)   declare a deadlock   end-if
end-while
```

The key point of (19) is why **L2** has deadlocks under the condition that **FPP** is not either **reducible** or **expansible**. This guarantee is provided by the first reduction operation of the algorithm. However, this paper does not focus on either how the reduction operation is implemented or why the operation guarantees a correct (19).

## 6. Conclusion

We have developed a software framework for examining synchronization communication deadlocks in MPI programs. The framework covers all methods presented in previous sections. The focus of this paper is not to address theory foundation and algorithm details of the framework but to demonstrate the process of executing the methods and algorithms in the framework because the volume of the whole theory is too big to be presented here and so many algorithms are used in our framework that they can not be covered here. In total this paper provides an overview of our work on static deadlock detection in MPI synchronization communication. Our methods are static so that all synchronization communication deadlocks can be found before real MPI-based programs go into running and losses incurred by deadlocks are reduced. From the perspective of synchronization communication deadlock detection, the method makes a greater process than known dynamic methods.